\documentclass[runningheads]{llncs}
\usepackage{graphicx}
\usepackage{multirow,multicol} %fjs
\usepackage{listings} % cite by fjs
\usepackage{algorithm} % cite by fjs
\usepackage{algorithmicx,algpseudocode} % cite by fjs
\usepackage{array} %fjs
\usepackage{bm}%fjs
\usepackage{amsfonts} % cite by fjs
\usepackage{url}%fjs 
\usepackage{epstopdf}%fjs eps->pdf
\usepackage{booktabs} %fjs
\usepackage[flushleft]{threeparttable} %fjs
\usepackage{makecell} %fjs
\usepackage{amsmath} %fjs 
\usepackage{mathtools} %fjs
\usepackage{color}

\begin{document}
\title{Weighted Concordance Index Loss-based Multimodal Survival Modeling for Radiation Encephalopathy Assessment in Nasopharyngeal Carcinoma Radiotherapy}
\titlerunning{Weighted CI Loss for NPC-REP}
% If the paper title is too long for the running head, you can set
% an abbreviated paper title here
%
\author{Jiansheng Fang \inst{1,2,3} \and Anwei Li \inst{2}  \and Pu-Yun OuYang \inst{4} \and Jiajian Li \inst{2} \and Jingwen Wang \inst{2} \and Hongbo Liu \inst{2} \and Fang-Yun Xie \inst{4} \and Jiang Liu \inst{3} \thanks{Corresponding author. Co-first authors: Jiansheng Fang, Anwei Li, Pu-Yun OuYang.} } 

%\thanks{Corresponding author: Jiang Liu (liuj@sustech.edu.cn).}}
%This work was supported in part by Guangdong Provincial Department of Education (2020ZDZX3043), and Shenzhen Natural Science Fund (JCYJ20200109140820699 and the Stable Support Plan Program 20200925174052004).

%\index{Fang Jiansheng}
%\index{Li Anwei}
%\index{OuYang Pu-Yun}
%\index{Li Jiajian}
%\index{Wang Jingwen}
%\index{Liu Hongbo}
%\index{Xie Fang-Yun}
%\index{Liu Jiang}

\authorrunning{Fang et al.}
% First names are abbreviated in the running head.
% If there are more than two authors, 'et al.' is used.
%
\institute{
School of computer science and technology, Harbin Institute of Technology, China %\email{11949039@mail.sustech.edu.cn}
\and CVTE Research, China % \email{lianwei@cvte.com}
%\and Department of Computer Science and Engineering, Southern University of Science and Technology, China
\and Research Institute of Trustworthy Autonomous Systems, Southern University of Science and Technology, China \email{liuj@sustech.edu.cn}
\and Department of Radiation Oncology, Sun Yat-sen University Cancer Center, China
% \email{ouyangpy@sysucc.org.cn}
}

%\author{Anonymous submission}
%\institute{Paper ID: 130}
%
\maketitle   % typeset the header of the contribution

\begin{abstract}
Radiation encephalopathy (REP) is the most common complication for nasopharyngeal carcinoma (NPC) radiotherapy. It is highly desirable to assist clinicians in optimizing the NPC radiotherapy regimen to reduce radiotherapy-induced temporal lobe injury (RTLI) according to the probability of REP onset. To the best of our knowledge, it is the first exploration of predicting radiotherapy-induced REP by jointly exploiting image and non-image data in NPC radiotherapy regimen. We cast REP prediction as a survival analysis task and evaluate the predictive accuracy in terms of the concordance index (CI). We design a deep multimodal survival network (MSN) with two feature extractors to learn discriminative features from multimodal data. One feature extractor imposes feature selection on non-image data, and the other learns visual features from images. Because the priorly balanced CI (BCI) loss function directly maximizing the CI is sensitive to uneven sampling per batch. Hence, we propose a novel weighted CI (WCI) loss function to leverage all REP samples effectively by assigning their different weights with a dual average operation. We further introduce a temperature hyper-parameter for our WCI to sharpen the risk difference of sample pairs to help model convergence. We extensively evaluate our WCI on a private dataset to demonstrate its favourability against its counterparts. The experimental results also show multimodal data of NPC radiotherapy can bring more gains for REP risk prediction.

\end{abstract}
\section{Introduction}
Radiotherapy is the standard radical treatment for nasopharyngeal carcinoma (NPC) and has considerably improved disease control and survival \cite{cavanna2006precuneus}. However,  NPC radiotherapy may induce radiation encephalopathy (REP), which makes the brain suffer irreversible damage during the incubation period \cite{tang2008relationship}. REP diagnosis remains challenging in clinical research due to its various clinical symptoms, insidious onset, long incubation period \cite{chen2019nasopharyngeal}. Currently, conventional magnetic resonance (MR) diagnosis can only discern REP at the irreversible stage \cite{chen2020altered}. Hence, it is highly desirable to assist clinicians in optimizing the radiotherapy regimen to reduce radiotherapy-induced temporal lobe injury (RTLI) according to the probability of REP onset. In this work, we study how to speculate REP risk in the pre-symptomatic stage by jointly exploiting the diagnosis and treatment data generated in NPC radiotherapy regimen, including computed tomography (CT) images, radiotherapy dose (RD) images, radiotherapy struct (RS) images, dose-volume histogram (DVH), and demographic characteristics. 

Recently, the data-driven approach has attracted a wide range of attention in the field of NPC. For example, many works for NPC segmentation have shifted from traditional hand-engineered models \cite{huang2015nasopharyngeal,huang2013region,zhou2006nasopharyngeal} to automatic deep learning models \cite{huang2019achieving,men2017deep}. Currently, there are a few studies that use data-driven approaches to predict the onset of REP. Zhao \textit{et al.} predict REP in NPC by analyzing the whole-brain resting-state functional connectivity density (FCD) of pre-symptomatic REP patients using multivariate pattern analysis (MVPA) \cite{zhao2021functional}. Zhang \textit{et al.} explore fractional amplitude of low-frequency fluctuation (fALFF) as an imaging biomarker for predicting or diagnosing REP in patients with NPC by using the support vector machine (SVM) \cite{zhang2021surface}. However, existing studies mainly consider the contribution of hand-engineered features computed from MR images to the REP prediction. The lesion region of REP stemming from NPC radiotherapy usually lie in the bilateral temporal lobe, and its severity is positively related to the dose \cite{wei2005nasopharyngeal}. Hence, according to the clinical evaluation mechanism, by casting REP risk prediction as a survival analysis task, we aim to build a deep multimodal survival network (MSN) to learn valuable and discriminative features from multiple data types generated during NPC radiotherapy.

The data types of NPC radiotherapy can be grouped into two modalities: image and non-image data. In MSN, we instinctively design two feature extractors to capture the discriminative information from multimodal data. One feature extractor explores the image data by convolution module, and the other mines the one-dimensional data by multi-layer perception (MLP) module. Given the concordance index (CI) \cite{harrell1996multivariable} is a standard performance metric for survival models, the existing balanced CI (BCI) loss function formulates the learning problem as directly maximizing the CI \cite{steck2008ranking}. However, BCI exhibits poor predictive accuracy due to its sensitivity to the sample pairing per batch. Hence, we propose a novel weighted CI (WCI) loss function with a temperature hyper-parameter $\tau$ to combat the sensitivity to uneven sampling and sharpen the relative risk of sample pairs, thus helping model convergence. We demonstrate our WCI advantage in terms of CI and 3-years area under the curve (AUC).

Our \textbf{\textit{contributions}} are: (a) To advance REP risk prediction to assess the rationality of NPC radiotherapy regimen, we propose a deep multimodal survival network (MSN) equipped with two feature extractors to learn discriminative features from image and non-image data. (b) We propose a novel WCI loss function with a temperature hyper-parameter for our MSN training to effectively leverage REP samples of each batch and help model convergence. (c) We confirm the benefits of our WCI with other loss functions used for survival models by extensively experimenting on a private dataset.

\section{Materials and Methods}

\subsection{NPC-REP Dataset Acquisition}
\begin{figure}[!t]
\centering
\includegraphics[width=\linewidth]{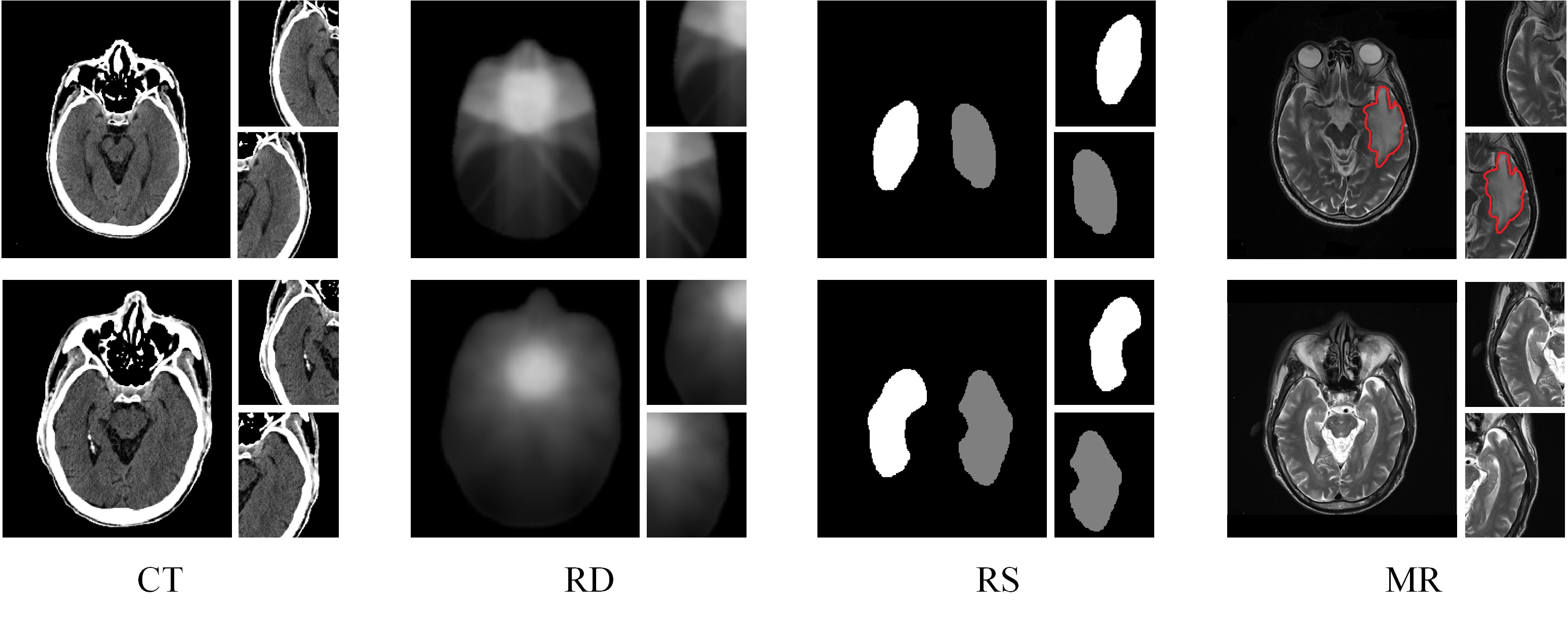} 
\caption{Schematic of image data in NPC-REP dataset. The upper row is a patient that confirms REP during follow-up, and the lower row is a patient without temporal lobe injury. The bilateral temporal lobe in MR images during follow-up place on column (MR). And by traceback, corresponding CT, RD, RS images generated in NPC radiotherapy lie in columns (CT), (RD), and (RS), respectively.}
\label{fig:img_data}
\end{figure}

We acquire eligible 4,816 patients from a cancer center. The eligibility of enrolling in the NPC-REP dataset is that diagnosis and treatment data in NPC radiotherapy and follow-up are traceable and recorded. We binarize a label for each patient according to whether the REP confirmation by MR image diagnosis during follow-up. If diagnosing out RTLI, the label of NPC patient is REP, otherwise non-REP. At the same time, we record the time intervals (in Months) between NPC radiotherapy and REP confirmation or the last follow-up of non-REP. Then we trace back the diagnosis and treatment data during NPC radiotherapy for each patient. RD images show the radiation dose distribution to the human body during radiation therapy. Clinicians sketch the tumor outline in closed curve coordinate form to yield the RS image. Then we apply the mask of RS images to generate the input ROIs of RD and CT images. As Fig \ref{fig:img_data} shows, we crop out the identical regions of interest (ROI) for CT images for diagnosis and RD images according to regions of the bilateral temporal lobe in RS images. Then we feed the three 3D ROIs into the multimodal learning network for extracting visual features. It is confirmed that the imaging dose delivered during NPC radiotherapy is positively related to REP risk \cite{wei2005nasopharyngeal}. Hence, in addition to considering image data, we introduce 18 features manually calculated from DVH, which relates radiation dose to tissue volume in radiation therapy planning \cite{drzymala1991dose}. Moreover, we also observe the affection of demographic information (age, gender), clinic stages (T/N/TNM), and treatment options (radio-chemotherapy and radiotherapy) to REP risk. The detailed statistics of demographic and clinical characteristics for the NPC-REP dataset are shown in Appendix A1.

\subsection{Multimodal Survival Modeling}

\begin{figure}[!t]
\centering
\includegraphics[width=\linewidth]{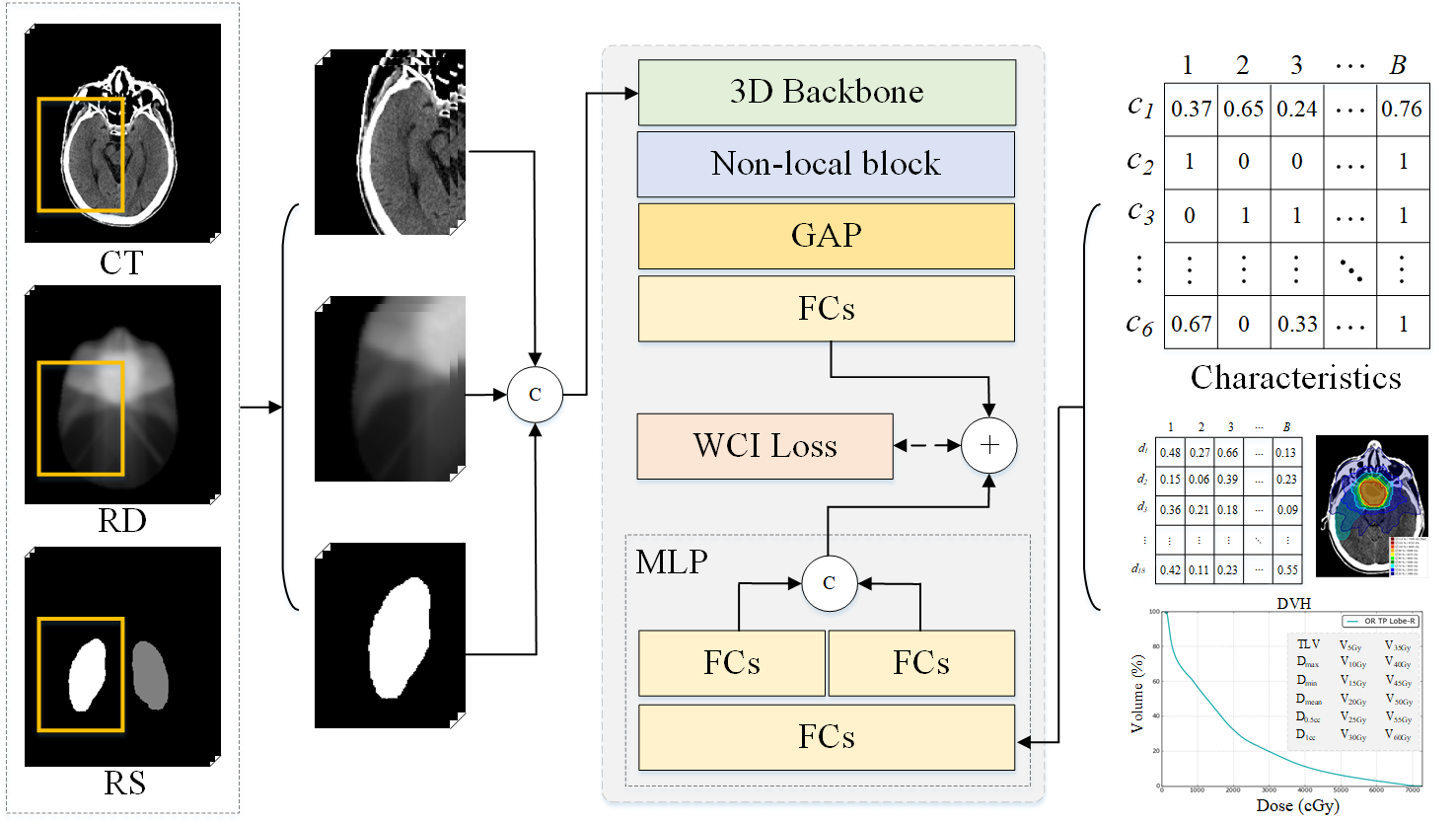} 
\caption{The architecture of our MSN with two feature extractors. Characteristics $c_{1,2,\dots 6}$ denote age, gender, T stage, N stage, TNM stage and treatment option. $d_{1,2,\dots 18}$ are manually calculated on DVH. FCs includes two fully-connected layers. GAP is a global average pooling layer. B is the batch size, and c indicates a concatenation operator.}
\label{fig:net_arch}
\end{figure}
We notice that most NPC patients with REP are diagnosed after radiotherapy for about 37 months. And it is almost impossible to occur REP when the follow-up time is less than two months and more than 73 months. Such a statistics observation about the onset time of REP illustrates that the risk probability of suffering REP is highly related to some factors in NPC radiotherapy. These factors cause the onset of REP in a relatively definite period. At the same time, we can cast REP assessment as a 3-years survival analysis task according to the average REP confirmation time intervals. Thus we build a deep multimodal survival network (MSN) to identify potential factors from the NPC-REP dataset, which consists of a binarized label, time intervals, image data (CT, RD, and RS images), and non-image data (24 features in total). As Fig \ref{fig:net_arch} shows, one feature extractor learns visual features of 3D ROIs intercepted from CT, RD, and RS images, and the other explores the contributions of non-image data to REP risk. The 24 features of non-image data are directly fed into the MLP module to output the REP risk $R_{nv}$. In the branch of visual feature extractor, we first apply convolution network as a backbone, followed by non-local attention capturing the global inter-dependencies \cite{wang2018non}. Next, we use a global average pooling (GAP) to merge feature maps and two fully-connected layers (FCs) to generate the REP risk $R_{v}$. After generating the two risk probabilities for two modality data, we further utilize our WCI loss function to train the network. The combination of the two predicted probabilities is defined as follows:
\begin{equation}
R = w_{nv}R_{nv} + w_{v}R_{v}
\label{equ:prob},
\end{equation}
where $R_{nv}, R_{v} \in [0,1]$ and $w_{nv}$ and $w_{v}$ are their corresponding weights. Both are initially set to 0.5. We can tune the two weights to observe the contribution of different modal data on assessing the REP risk.

\subsection{WCI Loss Function}
CI is a standard performance metric for survival models that corresponds to the probability that the model correctly orders a randomly chosen pair of patients in terms of event time \cite{chaudhary2018deep,wulczyn2020deep}. To calculate this indicator, we first pair all samples in the dataset of sample size $n$ with each other to get $n(n-1)\slash{2}$ sample pairs. Then, we filter out evaluation sample pairs from all sample pairs, as follows: 
\begin{equation}
\mathbb{E} \coloneqq \{(i,j) | (O_{i}=1 \wedge T_{j}\geq T_{i})\}
\label{equ:e},
\end{equation}
where $T_{i}$ and $T_{j}$ are the time intervals of $i^{th}$ and $j^{th}$ NPC patients respectively, and $O_{i}=1$ denotes that $i^{th}$ NPC patient is observed as REP. $\mathbb{E}$ is the set of sample pairs $(i,j)$ based on the ordering of survival times, \textit{i.e.}, the time intervals of $i^{th}$ NPC patient is shorter than $j^{th}$ NPC patient. Assuming the REP risks predicted by models for sample pairs $(i,j)$ are $R_{i}$ and $R_{j}$, we assert the prediction outcomes are correct if $R_{i}>R_{j}$ and $T_{i}<T_{j}$. We mark those sample pairs whose predicted results coincide with the actual as a set $\mathbb{E}_{t}$. The CI value is indicated as the number ratio of $\mathbb{E}_{t}\slash{\mathbb{E}}$. The higher CI explains away the better competence of survival models in predicting REP risks. 

Because the assumptions that proportional hazards are invariant over time and log-linear limit the Cox loss function (See Appendix A2) used for survival models, BCI loss function (See Appendix A3) directly formulates the optimizing problem of survival models as maximizing the CI. The Cox depends only on the ranks of the observed survival time rather than on their actual numerical value \cite{steck2008ranking}. Given the CI as a performance measure in survival analysis, BCI can effectively exploit the changes of proportional hazards with survival time. However, BCI is sensitive to uneven sampling, thus yielding poor predictive accuracy. Uneven sampling refers to the sample pairs meeting the constraints of Equation \ref{equ:e} in each batch failing to equally cover REP samples ($O_{i}=1$) which are the key learning objectives. BCI balances CI loss for all REP samples in each batch by adopting a global average operation, not paying more attention to those REP samples with fewer sample pairs. 

By referring to the weighted cross-entropy loss function, our WCI enhances the contributions of those REP samples with fewer sample pairs in each batch by a dual average operation, as follows:
\begin{equation}
\mathcal L_{wci} = \frac{1}{N_{O=1}} \sum_{i:O_{i}=1} (\frac{1}{N_{T_{j}\geq T_{i}}} \sum_{j:T_{j}\geq T_{i}} e^{-(R_{i}-R_{j})\slash{\tau}} )
\label{equ:wci},
\end{equation}
where $\tau$ is a temperature hyper-parameter \cite{he2020momentum,yi2019sampling} and initially set as 0.1. $N_{O=1}$ (outer) is the number of REP samples, and $N_{T_{j}\geq T_{i}}$ (inner) is the number of sample pairs of $i^{th}$ REP sample. If the relation of predicted risk probabilities is $R_{i}<R_{j}$ (wrong prediction), the CI loss is amplified, and if $R_{i}>R_{j}$ (correct prediction), the CI loss infinitely approximates to zero. After tuning the risk difference, we use an inner average operation $1\slash{N_{T_{j}\geq T_{i}}}$ and an outer average operation $1\slash{N_{O=1}}$ to prevent fluctuation in the optimization process caused by uneven sampling per batch. The outer average operation is balanced for all REP samples, while the inner average operation assigns different weights for REP samples according to their number of sample pairs. 

\section{Experiments}
\subsection{Implementation Details}
\textbf{Dataset.} We split the NPC patients in the NPC-REP dataset as a training set with 3,253 patients (264 REP and 2,989 non-REP), a validation set with 575 patients (43 REP and 532 non-REP), and a test set with 988 patients (61 REP and 927 non-REP). Moreover, we view left and right temporal lobes as independent samples during model training and inference. 

\textbf{Loss functions.} We experiment with four loss functions to demonstrate the gains of our WCI, including standard cross-entropy (CE) \cite{zadeh2020bias}, censored cross-entropy (CCE) \cite{wulczyn2020deep}, BCI \cite{steck2008ranking}, and Cox \cite{katzman2018deepsurv}. We adopt the survival time of 3-years (36 Months) for training and evaluation in terms of the time intervals in the NPC-REP dataset. For the CE, we define the label as 1 if REP confirmation and as 0 if the last follow-up time of non-REP is beyond 36 months. And the CCE extends CE used for classification models to train survival predictions with right-censored data by discretizing event time into intervals $(0,36], (36, \infty)$. The BCI maximizes a lower bound on CI, explaining the high CI-scores of proportional hazard models observed in practice. The Cox for every sample is a function of all samples in the training data and is usually used for fitting Cox proportional hazard models \cite{breslow1974covariance,cox1975partial}.

\textbf{Evaluation metrics.} CI is the most frequently used evaluation metric of survival models \cite{brentnall2018use,heller2016estimating}. It is defined as the ratio of correctly ordered (concordant) pairs to comparable pairs. Two samples $i$ and $j$ are comparable if the sample with lower observed time $T$ confirmed REP during follow-up, \textit{i.e.}, if $T_{j}>T_{i}$ and $O_{i}=1$, where $O_{i}=1$ is a binary event indicator. A comparable pair $(i,j)\in \mathbb{E}$ is concordant if the estimated risk $R$ by an assessment model is higher for NPC patients with lower REP confirmation time, \textit{i.e.}, $R_{i}>R_{j} \wedge T_{j}>T_{i} $, otherwise the pair is discordant. We also report the time-dependent AUC \cite{hung2010estimation,nuno2021censoring} for all loss functions, apart from CI.

\textbf{Optimizer.} Our MSN model is trained from scratch using the SGD optimizer for all compared loss functions. Specifically, the learning rate is decay scheduled by cosine annealing \cite{loshchilov2016sgdr} setting 2e-4 as an initial value and 1e-3 from 5th epochs. We set the batch size as a multiple of 8 for all tasks, \textit{i.e.}, 128 (8 GPUs with 16 input images per GPU). The parameters of networks are optimized in 60 epochs with a weight decay of 3e-5 and a momentum of 0.8.

\subsection{Results and Analysis}

\begin{figure}[!t]
\centering
\includegraphics[width=\linewidth]{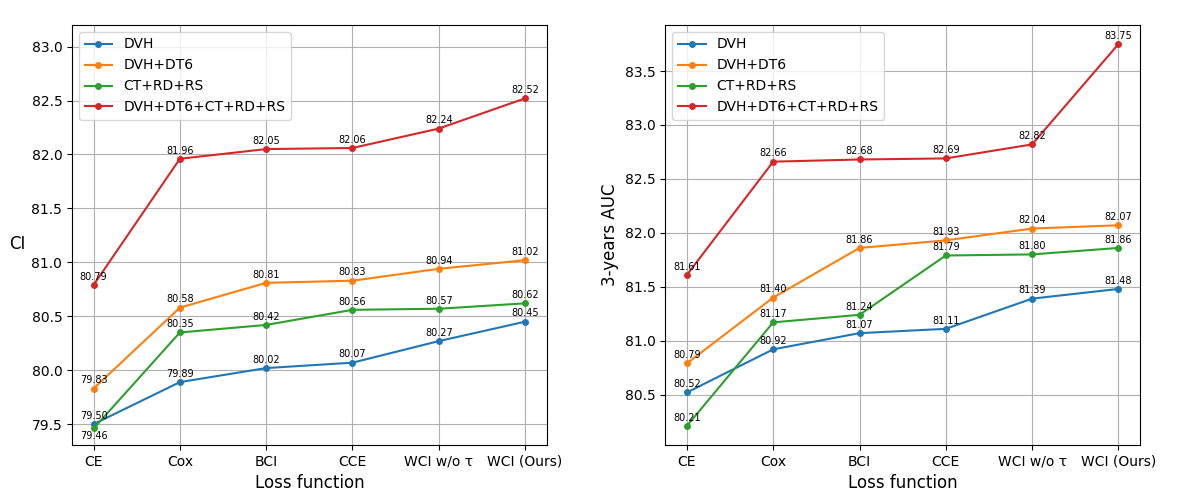} 
\caption{Comparison of different loss functions on our MSN for the NPC-REP dataset in terms of CI and 3-years AUC (in \%). DT6 covers age, gender, T/N/TNM stage, and treatment option, six features in total.}
\label{fig:cindex_auc}
\end{figure}
\textbf{Multimodal data Analysis.} 
The intention of acquiring treatment and diagnosis data in NPC radiotherapy launches on the confirmation that the imaging dose delivered during NPC radiotherapy is positively related to REP risk. Hence, we employ our MSN to jointly leverage non-image data (DVH+DT6) and images (CT+RD+RS) to predict REP risk. We report the CI and 3-years AUC in Fig \ref{fig:cindex_auc} for compared loss functions used for training our MSN. We can observe that combining DVH and DT6 (orange line) brings more gains than DVH (blue line) alone or CT+RD+RS (green line), and using all three data types (red line) achieves the best predictive accuracy. Such experimental results can demonstrate the contributions of multimodal data of NPC radiotherapy to the improvement of REP prediction and give supportive evidence for the above confirmation. Next, we analyze the contribution proportion of two modality data: images and non-image data. We conduct an ablation study for Equation \ref{equ:prob} to find the best weight ratio for two probabilities predicted by the two branches in our MSN. Table \ref{tab:risk_weight} shows the best CI (bold) achieved by setting the proportion of $w_{nv}$ and $w_{v}$ as 0.7:0.3. The more contributions of non-image data than images coincide with the reports of Fig \ref{fig:cindex_auc} in which non-image data (orange line) yields more gains than image data (green line).

\begin{table}[!t]
\caption{CI (in \%) of different loss functions with varying weight ratios.}
\label{tab:risk_weight}
\begin{center}
    \begin{tabular}{|c|c|c|c|c|c|c|}
    \toprule
    \hline
    $\bm{w_{nv}:w_{v}}$ & \textbf{CE}  & \textbf{Cox}  & \textbf{BCI} & \textbf{CCE} & \textbf{WCI w/o $\tau$} & \textbf{WCI (Ours)} \\
    \hline
    \textbf{$0.1 : 0.9$} & 79.54 & 79.33 & 79.47 & 79.65 & 80.25 & 80.53 \\
    \textbf{$0.3 : 0.7$} & 79.59 & 80.54 & 80.58 & 80.61 & 80.77 & 81.27 \\
    \textbf{$0.5 : 0.5$} & 80.01 & 81.47 & 81.66 & 81.54 & 81.66 & 81.89 \\
    \textbf{$0.7 : 0.3$} & \textbf{80.79}& \textbf{81.96} & \textbf{82.05} & \textbf{82.06} & 
    \textbf{82.24} &  \textbf{82.52} \\
    \textbf{$0.9:0.1$} & 80.65 & 81.32 & 81.42 & 81.50 & 81.53 & 81.62 \\
    \hline
    \bottomrule
    \end{tabular}
\end{center}
\end{table}

\textbf{Loss Function Analysis.}
After confirming the utility of multimodal data and analyzing the contributions of different modalities, we further discuss the benefits of our WCI according to the report in Fig \ref{fig:cindex_auc}. We interpret CI as the probability of correct ranking of REP occurring probability by casting predicting REP risk as a survival analysis task. Hence, CE used for learning classification models can not achieve sound accuracy without considering event times. Compared to CE, Cox can significantly improve the performance by exploiting the ranks of the observed survival time. On the other hand, Cox is beat by BCI due to ignoring the changes of proportional hazards with survival time. Although BCI enjoys the gain of directly maximizing CI,  its performance is slightly lower than CCE due to sensitivity to uneven sampling. Identical to BCI, CCE also considers the contributions of the proportional hazard of sample pairs in learning the survival model. Based on the comparable performance of BCI and CCE, WCI without $\tau$ can further obtain better performance by encouraging our MSN to effectively learn REP samples on the condition of uneven sampling per batch. By using $\tau$ help model convergence, our WCI achieves the highest CI (82.52) and 3-years AUC (83.75) on two modality data (DVH+DT6+CT+RD+RS). 
Although WCI is designed initially for the CI indicator used for evaluating survival models, the best performance on 3-years AUC can also prove the excellent generalization of our MSN trained by our WCI. We also conduct McNemar tests \cite{lachenbruch2014mcnemar} on CI between our WCI and other loss functions, including CE, Cox, BCI, CCE, and WCI w/o $\tau$. All comparisons in a p-value of less than 0.01 can support that our WCI brings significant improvement to CI.

\begin{table}[!t]
\caption{CI (in \%) of different data types with varying $\tau$ for our WCI loss function.}
\label{tab:temp_para}
\begin{center}
    \begin{tabular}{|c|c|c|c|c|c|}
    \toprule
    \hline
    \textbf{Data types} & $\bm{\tau = 10}$  & $\bm{\tau = 1}$  & $\bm{\tau = 0.1}$ & $\bm{\tau = 0.05}$ & $\bm{\tau = 0.02}$ \\
    \hline
    \textbf{DVH} & 80.23 & \underline{80.27} & \textbf{80.45} & 80.23 & 79.38 \\
    \textbf{DVH+DT6} & 80.85 & \underline{80.94} & \textbf{81.02} & 80.65 & 80.69 \\
    \textbf{CT+RD+RS} & 80.32 & \underline{80.57} & \textbf{80.62} & 80.29 & 79.53 \\
    \textbf{DVH+DT6+CT+RD+RS} & 80.93 & \underline{82.24} & \textbf{82.52} & 81.16 & 80.76 \\
    \hline
    \bottomrule
    \end{tabular}
\end{center}
\end{table}

\textbf{Ablation studies of $\tau$.}
By casting REP assessment as a survival analysis task, our WCI exhibits manifest advantage over other loss functions by introducing the dual average operation and temperature hyper-parameter $\tau$. Apart from the dual average operation preventing our MSN from being sensitive to uneven sampling, we also utilize $\tau$ to sharpen the risk difference of sample pairs to help model convergence. We conduct an ablation study for $\tau$ to observe the heating and cooling effect of tuning the CI loss for our WCI. Table \ref{tab:temp_para} shows that the best value of $\tau$ used to train our MSN is 0.1 (bold). And when $\tau=1.0$, our WCI without imposing the tuning on CI achieves the second-highest CI of 82.24 (underline) over two modality data (DVH+DT6+CT+RD+RS). When $\tau>1.0$, it occurs the opposite effect. At the same time, the adjustment force should not be too large from the performance of setting $\tau=0.05$ or $\tau=0.02$. The ablation studies in Table \ref{tab:temp_para} confirm that $\tau$ can help model convergence by sharpening the risk difference. Theoretically, the significant risk difference helps improve the predicted certainty probability, \textit{i.e.}, reduces the entropy of probability distribution. Benefiting from the insensitive to uneven sampling and sharpening risk difference, our WCI exhibits more stability and convergence than BCI by comparing standard deviations of batch loss (See Appendix A4).

\section{Conclusions}
We present a novel WCI loss function to effectively exploit REP samples per batch by a dual average operation and help model convergence by sharpening risk difference with a temperature hyper-parameter. It is the first exploration of jointly leveraging multimodal data to predict the probability of REP onset to help optimize the NPC radiotherapy regimen. We acquire a private dataset and experiment on it to demonstrate the favorability of our WCI by comparing it to four popular loss functions used for survival models. We also confirm the contributions of multimodal data to REP risk prediction. 

%
% ---- Bibliography ----
%
% BibTeX users should specify bibliography style 'splncs04'.
% References will then be sorted and formatted in the correct style.
%
\bibliographystyle{splncs04}
\bibliography{cindex}

\end{document}